\documentclass
[superscriptaddress,secnumarabic,amssymb,amsmath,nobibnotes,aps,prd,showkeys,showpacs,nofootinbib,onecolumn,12pt]{revtex4}%
\usepackage{graphicx}
\usepackage{epsf}
\usepackage{bm}
\usepackage{amsmath}
\usepackage{amsfonts}
\usepackage{amssymb}%
\providecommand{\U}[1]{\protect\rule{.1in}{.1in}}

\newcommand{\be}{\begin{equation}}
\newcommand{\ee}{\end{equation}}

\newcommand{\mincir}{\raise
-3.truept\hbox{\rlap{\hbox{$\sim$}}\raise4.truept\hbox{$<$}\ }}
\newcommand{\magcir}{\raise
-3.truept\hbox{\rlap{\hbox{$\sim$}}\raise4.truept\hbox{$>$}\ }}

\begin{document}
\title{Unified dark energy from Chiral-Quintom model with a mixed potential in
Friedmann--Lema\^{\i}tre--Robertson--Walker cosmology}
\author{Andronikos Paliathanasis}
\email{anpaliat@phys.uoa.gr}
\affiliation{Institute of Systems Science, Durban University of Technology, Durban 4000,
Republic of South Africa}
\affiliation{Departamento de Matem\'{a}ticas, Universidad Cat\'{o}lica del Norte, Avda.
Angamos 0610, Casilla 1280 Antofagasta, Chile}

\begin{abstract}
For a spatially Friedmann-Lema\^{\i}tre-Robertson-Walker cosmology, we propose
a multi-scalar field gravitational model. Specifically, we consider a
two-scalar field cosmological model in which the kinetic components of the
scalar fields establish a two-dimensional sphere of Lorentzian signature. For
our Chiral-Quintom model we choose a mixed potential term $V\left(  \phi
,\psi\right)  =V_{0}e^{\lambda\phi}+U_{0}e^{\kappa\phi}\psi^{\frac{1}{\sigma}%
}$ and we investigate the asymptotic limits of the cosmological parameters.
This model for $U_{0}=0$, provides a generalization of the hyperbolic
inflation where the equation of the state parameter can cross the phantom
divide line. When $U_{0}\neq0$ we observe that this cosmological model
exhibits asymptotic solutions that encompass accelerated universes, big rip
singularities, and dust-like solutions. Hence, this multi-scalar field model
it can be regarded as as a dark energy unify model which describes a variety
of asymptotic cosmological scenarios.

\end{abstract}
\keywords{Chiral-Quintom model; Dark Energy; Unified dark energy}\date{\today}
\maketitle

\section{Introduction}

Quintom cosmology falls within the family of multi-scalar fields gravitational
theories. \cite{q1,q2,q3}. In this theory the cosmological fluid comprises two
scalar fields that are minimally coupled to gravity. One of these scalar
fields is the quintessence scalar \cite{q4,q5} and the other is the phantom
scalar field \cite{q6,q7} which possess distinct properties.

The quintessence is one of the first introduced dynamical models to describe
the expansion of the universe. The quintessence field is characterized as
inflaton during the early-time acceleration phase of the universe
\cite{q8,q9}, while in the\ present time acceleration, the quintessence
attributes the dark energy components of the cosmological fluid \cite{q10,q11}%
. Quintessence scalar field has been used to unify the dark matter and the
dark energy \cite{q12}. The quintessence scalar field adheres to the null
energy condition, the weak energy condition, and the dominant energy
condition, while allowing for the violation of the strong energy condition in
order to provide acceleration.

In contrast, in the case of a phantom field, the equation of state parameter
is not constrained by a lower boundary. This implies that it has the ability
to cross the phantom divide line, allowing for the energy density to become
negative. Consequently, the phantom scalar field violates all the energy
conditions, which means that the equation of state parameter can cross the
phantom divide line. It worth to mention that for the phantom scalar field
model, the equation of state parameter can cross the phantom divide line only
once. The results from the statistical analysis of the cosmological
observations do not exclude the equation of state parameter for the
cosmological fluid to take values smaller than that of minus one \cite{qq1}.
This cosmological fluid can be the source for the so-called Big-Rip
\cite{q14}. The detailed analysis of the cosmological dynamics for the phantom
scalar fields has shown that for an unbounded scalar field potential the
late-time attractor can describe a super-exponential universe which leads to a
Big-Rip or other kind of sudden singularities; however for a bounded scalar
field potential function the de Sitter spacetime is a future attractor
\cite{q15}.

The quintom cosmology it is a multi-scalar field model that involves the
dynamics of two scalar fields, namely quintessence and phantom fields. It was
proposed to overcome the limitations imposed by single scalar field models on
the equation of state parameter Indeed, in the quintom cosmology, there exist
epochs provided by the cosmological dynamics where the quintessence field
dominates; or the phantom field dominates and there exist solutions where the
two scalar field contributes in the cosmic fluid. As a result it is possible
the equation of state parameter for the quintom model to cross more than once
the phantom divide line \cite{q16}. This is due to the presence of both
quintessence and phantom fields in the cosmic dynamics. A detailed analysis of
the evolution for the cosmological parameters in quintom model for various
potential functions is presented in \cite{q17,q18}. It was found that for
various forms of the potential functions there can be periods in the provided
cosmological history where one field dominates, followed by the an epoch where
the other field dominates, as also there exist asymptotic solutions where both
fields provide in the cosmic fluid. Recently in \cite{q18a} it was introduced
a quintom model with mixed potential term. This model has been used to fit the
cosmological observations and it was found that it can describe well the
recent cosmological data. On the other hand, in \cite{q25} the quintom model
has been proposed to investigate if it solves Hubble tension problem. Because
of the importancy of the quintom model, there are a plethora of studies in the
literature, we refer the reader in \cite{q19,q20,q21,q22,q23,q24,q26} and
references therein. There are various extensions and generalizations of the
quintm model in modified theories of gravity, for instance in scalar-tensor
theory \cite{q18ab,q18b}, in Galilleon cosmology \cite{q18c}, in
scalar-torsion theory \cite{q18d}, in Gauss-Bonnet theory \cite{q18e}. 

The Chiral model, a cosmological fluid characterized by two scalar fields, has
garnered significant attention and undergone extensive study in recent years.
In the context of General Relativity, the Lagrangian for the matter source
describes two scalar fields where interaction exists between the two scalar
fields in the components. Chiral model is part of the family of the non-linear
sigma model \cite{sigm0}, where the two-scalar fields are defined on a
two-dimensional sphere \cite{atr6,atr7}. Previous studies of Chiral theory
have shown that two acceleration phases for the universe are provided by the
model. The slow-roll inflation is recovered in the limit where the model
reduces to quintessence, and the second acceleration phase is known as the
hyperbolic inflation \cite{vr91,pp} where the two scalar fields contribute in
the cosmic evolution. Because of the existence of the second scalar field in
the hyperbolic inflation there are some characteristic differences with the
slow-roll inflation. Notably, the initial conditions at the onset and the end
of inflation can be different, and the curvature perturbations being
contingent upon the number of e-folds \cite{ch1}. Furthermore, as it has been
shown in \cite{ch2} non-Gaussianities in the power spectrum are provided by
the\ Chiral theory. Although the field equations of Chiral model are
non-linear there are various studies where analytic and closed-form solutions
are presented \cite{ex1,ex2}.

In the case of the exponential potential, an extensive examination of the
phase-space encompassing the physical variables and the identification of
asymptotic solutions for Chiral cosmology are meticulously outlined in
\cite{cher1}. For a more generic potential function we refer the reader in
\cite{dn1}. The Chiral model with a mixed potential term can be seen as a
unified dark model which means that it can describe the fluid components which
contribute to the dark sector of the universe. In the very early universe in
Chiral theory there exists a mechanism based on quantum transitions where the
effective cosmological fluid can have an equation of state parameter which can
cross the phantom divide line and provide a rapid expansion of the universe
\cite{dn2}. Extensions of the Chiral model with more \ that two scalar fields
have been considered before in \cite{dn5,dn6}. Furthermore, hyperbolic
inflation in the presence of spatial curvature is discussed in \cite{dn7}, it
was found that the hyperbolic inflationary solutions can solve the flatness
problem and describe acceleration for both open and closed models.

The Chiral cosmological model is characterized by a fundamental property
wherein the effective energy density of the cosmological fluid remains
positive by definition \cite{atr6}. Furthermore, the effective equation of
state parameter within the Chiral cosmological model is subject to a lower
limit that precisely matches the value of the cosmological constant
\cite{atr6}. Inspired by the quintom model in \cite{dn8} two families of
Chiral-Quintom models were proposed. From the analysis of the asymptotic it
was found that the Chiral-Quintom model has similar properties to the quintom
model, while the hyperbolic inflation is supported \cite{dn9}. Moreover, the
dynamical evolution of the physical parameters in the Chiral-Quintom theory in
the presence of curvature\ was recently investigated in \cite{dn10}.
In\ Chiral-Quintom theory, the interaction for the scalar fields is the same
as in the Chiral model, but what changes is the signature of the
two-dimensional manifold which defines the dynamics for the scalar fields.
There are two possible models \cite{dn8}; however from the analysis of the
dynamics \cite{dn9} it was found that one of this provides an cosmological
history which can explain the major eras of the cosmological evolution.

The objective of this study is to examine the dynamics of the Chiral-Quintom
cosmological model with a mixed potential term. Our investigation aims to
assess the viability of utilizing the Chiral-Quintom model as a simplified
framework for unifying the components within the dark sector of the universe.
In particular we extend the analysis presented in \cite{dn1} for the case
where the second scalar field is a phantom field. This cosmological model
holds the potential to establish a connection between various epochs of cosmic
evolution, elucidating phenomena such as inflation, the matter era, and the
late-time acceleration phase. The plan of this study is outlined as follows.

In\ Section \ref{sec2} we discuss the basic definitions of the Chiral-Quintom
model of our consideration. Additionally, we present the field equations
specifically in the context of a spatially flat
Friedmann-Lema\^{\i}tre-Robertson-Walker (FLRW) cosmology. In\ Section
\ref{sec3} we reformulate the field equations using dimensionless variables
for a more convenient and comprehensive analysis. The main results of this
analysis are presented in Sections \ref{sec4a} and \ref{sec4}. In these
sections, we focus on investigating the asymptotic limits of the field
equations for two different mixed potential functions. We analyze the
evolution of the physical parameters and thoroughly examine the stability
properties of the asymptotic solutions at both the stationary points in the
finite regime and those in the infinity regime. Finally, in Section \ref{sec5}
we draw our conclusions.

\section{Chiral-Quintom Cosmology with Mixed Potential}

\label{sec2}

The Action Integral of the Chiral model is defined within the framework of
General Relativity, that is, \cite{dn1}
\begin{equation}
S=S_{EH}+S_{Chiral}, \label{ac.01}%
\end{equation}
in which $S_{EH}$ is the Einstein-Hiblert Action Integral%
\begin{equation}
S_{EH}=\int\sqrt{-g}dx^{4}R,
\end{equation}
and $S_{Chiral}$ attributes the dynamical terms of the two-scalar fields%
\begin{equation}
S_{Chiral}=\int\sqrt{-g}dx^{4}\left(  -\frac{1}{2}g^{\mu\nu}H_{AB}\left(
\Phi^{C}\right)  \nabla_{\mu}\Phi^{A}\nabla_{\nu}\Phi^{B}-V\left(  \Phi
^{C}\right)  \right)  ,
\end{equation}
with $\Phi^{A}=\left(  \phi\left(  x^{\mu}\right)  ,\psi\left(  x^{\mu
}\right)  \right)  ^{T},~V\left(  \Phi^{C}\right)  $ is the potential function
and $H_{AB}\left(  \Phi^{C}\right)  $ is a second-rank tensor which defines
the space where the scalar fields are defined.

For the Action Integral (\ref{ac.01}) the Einstein field equations are%
\begin{equation}
G_{\mu\nu}=H_{AB}\left(  \Phi^{C}\right)  \nabla_{\mu}\Phi^{A}\nabla_{\nu}%
\Phi^{B}-g_{\mu\nu}\left(  \frac{1}{2}g^{\mu\nu}H_{AB}\left(  \Phi^{C}\right)
\nabla_{\mu}\Phi^{A}\nabla_{\nu}\Phi^{B}+V\left(  \Phi^{C}\right)  \right)  ,
\end{equation}
while for the scalar fields the equations of motion read%
\begin{equation}
g^{\mu\nu}\left(  \nabla_{\mu}\left(  H_{~B}^{A}\left(  \Phi^{C}\right)
\nabla_{\nu}\Phi^{B}\right)  \right)  +H_{~B}^{A}\left(  \Phi^{C}\right)
\frac{\partial V\left(  \Phi^{C}\right)  }{\partial\Phi^{B}}=0.
\end{equation}

In Chiral model $H_{AB}\left(  \Phi^{C}\right)  $ is considered to be
described by the second rank tensor \cite{vr91}%
\begin{equation}
H_{AB}\left(  \Phi^{C}\right)  =diag\left(  1,e^{\kappa\phi}\right)  ,
\end{equation}
with signature $\left(  +,+\right)  $.

However, in the Chiral-Quintom theory of our consideration we assume the
signature to be $\left(  +,-\right)  $, hence we set \cite{dn8}%
\begin{equation}
H_{AB}\left(  \Phi^{C}\right)  =diag\left(  1,-e^{\kappa\phi}\right)  .
\label{ac.02}%
\end{equation}

\subsection{FLRW Cosmology}

On large scales, the physical structure of the universe is represented by the
spatially flat FLRW geometry, given by the line-element%
\begin{equation}
ds^{2}=-dt^{2}+a^{2}\left(  t\right)  \left(  dx^{2}+dy^{2}+dz^{2}\right)  .
\label{ac.06}%
\end{equation}
where function $a\left(  t\right)  $ is scale factor. The Hubble function is
determined $H\left(  t\right)  =\frac{\dot{a}}{a}$. Additionally, the
expansion rate for a comoving observer, $u^{\mu}=\delta_{t}^{\mu}$ is defined
as $\theta=u_{;\mu}^{\mu}$, that is, $\theta\left(  t\right)  =3H\left(
t\right)  $.

The FLRW spacetime (\ref{ac.06}) possesses a sixth-dimensional Killing
algebra. Consequently, if we assume that the scalar fields inherit the
symmetries of the background space, we arrive at the conclusion that the
scalar fields $\phi\left(  x^{\mu}\right)  =\phi\left(  t\right)  $,
$\psi\left(  x^{\mu}\right)  =\psi\left(  t\right)  $.

Following the analysis presented in \cite{dn1}, we adopt a mixed scalar field
potential%
\begin{equation}
V\left(  \Phi^{C}\right)  =V\left(  \phi\right)  +e^{\kappa\phi}U\left(
\psi\right)  .
\end{equation}

Thus, for the latter potential function and the second rank tensor
(\ref{ac.02}) we can derive the Friedmann's equations as follows:%
\begin{align}
3H^{2}  &  =\frac{1}{2}\dot{\phi}^{2}-\frac{1}{2}e^{\kappa\phi}\dot{\psi}%
^{2}+V\left(  \phi\right)  +e^{\kappa\phi}U\left(  \psi\right)  ,\label{ss.05}%
\\
-\left(  2\dot{H}+3H^{2}\right)   &  =\frac{1}{2}\dot{\phi}^{2}-\frac{1}%
{2}e^{\kappa\phi}\dot{\psi}^{2}-\left(  V\left(  \phi\right)  +e^{\kappa\phi
}U\left(  \psi\right)  \right)  . \label{ss.06}%
\end{align}

Furthermore, the scalar fields obey the system of Klein-Gordon equations:%
\begin{equation}
\left(  \ddot{\phi}+3H\dot{\phi}\right)  +\frac{1}{2}\kappa e^{\kappa\phi}%
\dot{\psi}^{2}+V_{,\phi}+\kappa e^{\kappa\phi}U\left(  \psi\right)  =0~,
\label{ss.07}%
\end{equation}%
\begin{equation}
\ddot{\psi}+3H\dot{\psi}+\kappa\dot{\phi}\dot{\psi}+U_{,\psi}=0~,
\label{ss.08}%
\end{equation}

From (\ref{ss.05})-(\ref{ss.06}) we can define the effective energy density
and pressure components%
\begin{equation}
\rho_{\phi}=\frac{1}{2}\dot{\phi}^{2}+V\left(  \phi\right)  ~,~p_{\phi}%
=\frac{1}{2}\dot{\phi}^{2}-V\left(  \phi\right)  ,
\end{equation}%
\begin{equation}
\rho_{\psi}=\left(  -\frac{1}{2}\dot{\psi}^{2}+U\left(  \psi\right)  \right)
e^{\kappa\phi}~,~p_{\psi}=\left(  -\frac{1}{2}\dot{\psi}^{2}-U\left(
\psi\right)  \right)  e^{\kappa\phi},
\end{equation}

With the use of the fluid components the filed equations (\ref{ss.05}),
(\ref{ss.06}) can be written in the traditional form%
\[
G_{\mu\nu}=T_{\mu\nu}^{\left(  \phi\right)  }+T_{\mu\nu}^{\left(  \psi\right)
}%
\]
where now~$T_{\mu\nu}^{\left(  \phi\right)  }$ and $T_{\mu\nu}^{\left(
\psi\right)  }$ attributes the components of the two interact scalar fields,
that is,
\begin{align}
T_{\mu\nu}^{\left(  \phi\right)  }  &  =\left(  \rho_{\phi}+p_{\phi}\right)
u_{\mu}u_{\nu}+p_{\phi}g_{\mu\nu},\\
T_{\mu\nu}^{\left(  \psi\right)  }  &  =\left(  \rho_{\psi}+p_{\psi}\right)
u_{\mu}u_{\nu}+p_{\psi}g_{\mu\nu},
\end{align}

Moreover, the continuous equation is $\left(  T^{e_{ff~~\mu\nu}}\right)
_{;\nu}=0$, or equivalent $\left(  T^{\left(  \phi\right)  \mu\nu}+T^{\left(
\psi\right)  \mu\nu}\right)  _{;\nu}=0$. Since the two scalar fields interact,
we can write the continuous equation as $\left(  T^{\left(  \phi\right)
\mu\nu}\right)  _{;\nu}=Q$, $\left(  T^{\left(  \psi\right)  \mu\nu}\right)
_{;\nu}=-Q$, which are the two equations of motion for the scalar fields
$\phi$ and $\psi$, equations (\ref{ss.07}) and (\ref{ss.08}) if we select
$Q=\kappa\dot{\phi}\dot{\psi}$.

For the two scalar fields we can define the equation of state parameters as%
\begin{equation}
w_{\phi}=\frac{\frac{1}{2}\dot{\phi}^{2}-V\left(  \phi\right)  }{\frac{1}%
{2}\dot{\phi}^{2}+V\left(  \phi\right)  }~,~w_{\psi}=\frac{-\frac{1}{2}%
\dot{\psi}^{2}-U\left(  \psi\right)  }{-\frac{1}{2}\dot{\psi}^{2}+U\left(
\psi\right)  }.
\end{equation}
Thus, $\left\vert w_{\phi}\right\vert \leq1$, while $w_{\psi}$ can take values
smaller than minus one. The limit where $U\left(  \psi\right)  =0$, was
investigated in \cite{dn8} where in this piece of study the second scalar
field $\psi$ describes a stiff fluid with $w_{\psi}=1$. In our consideration
$w_{\psi}$ is a dynamical variable, since $U\left(  \psi\right)  $ is assumed
to be a nonzero variable.

Finally, for the effective cosmological fluid we find%
\begin{equation}
w_{tot}=\frac{\frac{1}{2}\dot{\phi}^{2}-V\left(  \phi\right)  +e^{\kappa\phi
}\left(  -\frac{1}{2}\dot{\psi}^{2}+U\left(  \psi\right)  \right)  }{\frac
{1}{2}\dot{\phi}^{2}+V\left(  \phi\right)  +e^{\kappa\phi}\left(  -\frac{1}%
{2}\dot{\psi}^{2}+U\left(  \psi\right)  \right)  }.
\end{equation}

\section{Autonomous dynamical system}

\label{sec3}

We proceed our study by defining new variables $\left(  x,y,z,u,\lambda
,\mu\right)  $ in the so-called $H$-normalization approach \cite{dn1}%
\begin{equation}
\dot{\phi}=\sqrt{6}xH~,~V\left(  \phi\right)  =3y^{2}H^{2}~,~\dot{\psi}%
=\sqrt{6}e^{-\frac{\kappa}{2}\phi}zH~, \label{ch.01}%
\end{equation}%
\begin{equation}
~U\left(  \psi\right)  =3e^{-\kappa\phi}u^{2}H^{2}~,~V_{,\phi}=\lambda
V~,~U_{,\psi}=e^{\frac{\kappa}{2}\phi}\mu U~.~
\end{equation}
With the application of the latter dimensionless variables the field equations
are written in the equivalent form of a system of first-order ordinary
differential equations.

Friedmann's equation (\ref{ss.05}) provides the algebraic equation
\begin{equation}
1-x^{2}-y^{2}+z^{2}-u^{2}=0. \label{ch.00}%
\end{equation}

Furthermore, for the field equations (\ref{ss.06})-(\ref{ss.08}) it follows%

\begin{align}
\frac{dx}{d\tau}  &  =\frac{3}{2}x\left(  x^{2}-\left(  1+u^{2}+y^{2}%
+z^{2}\right)  \right)  -\frac{\sqrt{6}}{2}\left(  \lambda y^{2}+\kappa\left(
u^{2}+z^{2}\right)  \right)  ,\label{ch.02}\\
\frac{dy}{d\tau}  &  =\frac{3}{2}y\left(  1+x^{2}-z^{2}-y^{2}-u^{2}\right)
+\frac{\sqrt{6}}{2}\lambda xy,\label{ch.03}\\
\frac{dz}{d\tau}  &  =-\frac{3}{2}z\left(  z^{2}+y^{2}-x^{2}+u^{2}-1\right)
-\frac{\sqrt{6}}{2}\left(  \kappa xz-\mu u^{2}\right)  ,\label{ch.04}\\
\frac{du}{d\tau}  &  =\frac{3}{2}u\left(  1+x^{2}-z^{2}-y^{2}-u^{2}\right)
+\frac{\sqrt{6}}{2}u\left(  \kappa x+\mu z\right)  ,\label{ch.05}\\
\frac{d\mu}{d\tau}  &  =\sqrt{\frac{3}{2}}\mu\left(  2\mu z\left(  \bar
{\Gamma}\left(  \mu,\lambda\right)  -1\right)  -\kappa x\right)
,\label{ch.06}\\
\frac{d\lambda}{d\tau}  &  =\sqrt{6}\lambda^{2}x\left(  \Gamma\left(
\lambda\right)  -1\right)  , \label{ch.07}%
\end{align}
where now the new independent variable is $\tau=\ln a$, and functions
$\Gamma\left(  \lambda\right)  ,~\bar{\Gamma}\left(  \mu,\lambda\right)  $ are%
\begin{equation}
\Gamma\left(  \lambda\right)  =\frac{V_{,\phi\phi}V}{\left(  V_{,\phi}\right)
^{2}}~,~\bar{\Gamma}\left(  \mu,\lambda\right)  =\frac{U_{,\psi\psi}U}{\left(
U_{,\psi}\right)  ^{2}}, \label{ch.09}%
\end{equation}

At this point we remark that on the constant surface $u=0$, where $U\left(
\psi\right)  =0$, the latter dynamical system is reduced to the special case
studied before in \cite{dn8}. However, in our study, we introduce a non-zero
potential $U\left(  \psi\right)  $ which significantly impacts the dynamics
and introduces new asymptotic solutions in the cosmological model.
\begin{equation}
\Omega_{\phi}=x^{2}+y^{2}~,~\Omega_{\psi}=-z^{2}+u^{2},
\end{equation}
where $\Omega_{\phi}$ and $\Omega_{\psi}$ are the energy densities for the two
scalar fields. \ 

Consequently, the equation of state parameters corresponding to the fields
$\phi$ and $\psi$ are as follows
\begin{equation}
w_{\phi}=-1+\frac{2x^{2}}{x^{2}+y^{2}}~,~w_{\psi}=-1-\frac{2z^{2}}{u^{2}%
-z^{2}}.
\end{equation}

On the contrary, for the effective cosmological fluid, the equation of state
parameter reads%
\begin{equation}
w_{tot}=-1-\frac{2}{3}\frac{\dot{H}}{H^{2}}=x^{2}-z^{2}-y^{2}-u^{2}%
\end{equation}

For a general potential function $V\left(  \phi,\psi\right)  $, the dynamical
system  (\ref{ch.02})-(\ref{ch.07}) has dimension six. Nevertheless for the
exponential potential function $V\left(  \phi\right)  =V_{0}e^{\lambda\phi}$,
we derive $\Gamma\left(  \lambda\right)  =\lambda\,$, that is, $\lambda$ is
always a constant parameter and the dimension of the system  (\ref{ch.02}%
)-(\ref{ch.07}) is reduced by one. Furthermore, for this potential, with the
use of the algebraic equation (\ref{ch.00}) we end with a four-dimensional
dynamical system. Given its connection with hyperbolic inflation, we focus our
subsequent analysis on the exponential potential $V\left(  \phi\right)
=V_{0}e^{\lambda\phi}$ . As for the second scalar field, we adopt the
power-law potential $U\left(  \psi\right)  =U_{0}\psi^{\frac{1}{\sigma}}$. It
is worth noting that from this potential function, we derive $\bar{\Gamma
}\left(  \mu,\lambda\right)  =1-\sigma,~$where $\sigma$ is a constant$~$%
parameter \cite{dn1}. The power-law potential $U\left(  \psi\right)
=U_{0}\psi^{\frac{1}{\sigma}}~$is of important interest, because on the
surface where the scalar field $\phi$ is constant, that is, $\phi~$does not
contribute in the universe, then the de Sitter solution is provided by the
dynamical terms of the second scalar field $\psi$. \ Therefore, we proceed our
analysis with\ the selection $V\left(  \phi,\psi\right)  =V_{0}e^{\lambda\phi
}+U_{0}e^{\kappa\phi}\psi^{\frac{1}{\sigma}}$. The limit where $V_{0}=0$, it
is examined individually.

The variables $y$ and $u$ are strictly positive, indicated by the conditions
$y\geq0$ and $u\geq0.$ On the other hand, the variables $x\,$and $z$ can
assume any real number value within their respective ranges. It is important
to note that in Chiral theory, all parameters are constrained to reside on the
surface of a four-dimensional sphere. However, in this particular case, of the
Chiral-Quintom theory, such restrictions do not apply. Consequently, a
thorough investigation of the asymptotic behavior at infinity becomes necessary.

We compute the stationary/critical points of the dynamical system
(\ref{ch.02})-(\ref{ch.06}). Each stationary point corresponds to a specific
asymptotic solution governing the background geometry. At these stationary
points, it becomes possible to determine the values of the physical parameters
and reconstruct the associated asymptotic solutions.

For $w_{tot}\neq-1\,,$ the asymptotic solution describes a scaling solution
with $a\left(  t\right)  =a_{0}t^{\frac{2}{3\left(  1+w_{tot}\right)  }}$, and
acceleration is occurred for $w_{tot}<-\frac{1}{3}$. Besides $w_{tot}=-1$, the
asymptotic solution describes a de Sitter universe with exponential scale
factor $a\left(  t\right)  =a_{0}e^{H_{0}t}$, where the effective cosmological
fluid is described by the cosmological constant.

\section{Asymptotic solutions for potential $V\left(  \phi,\psi\right)
=V_{0}e^{\lambda\phi}+U_{0}e^{\kappa\phi}\psi^{\frac{1}{\sigma}}$}

\label{sec4}

Let us now consider the scenario where the mixed potential function takes the
form $V\left(  \phi,\psi\right)  =V_{0}e^{\lambda\phi}+U_{0}e^{\kappa\phi}%
\psi^{\frac{1}{\sigma}}$.

\subsection{Stationary points at the finite regime}

\begin{figure}[ptb]
\centering\includegraphics[width=1\textwidth]{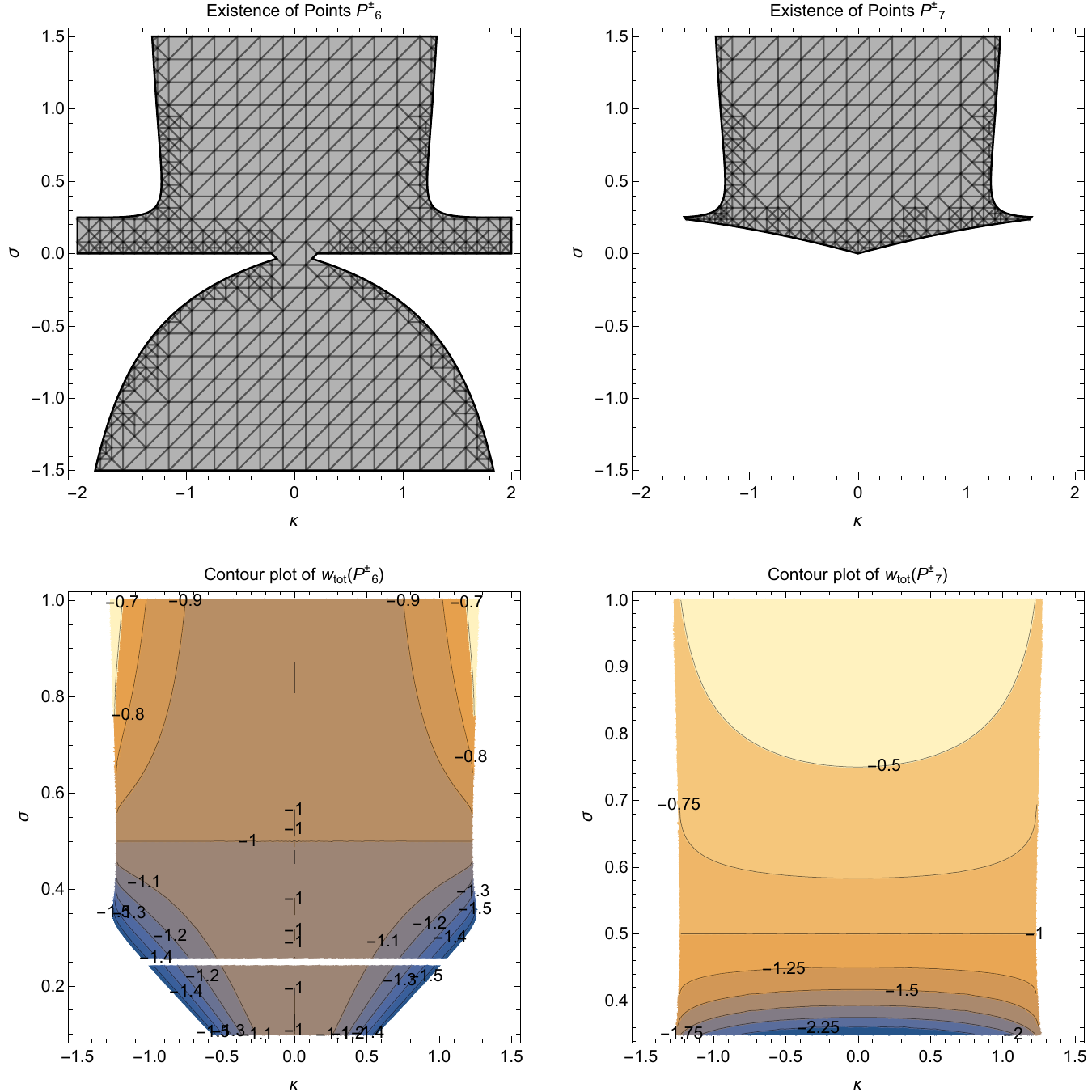} \caption{First Column:
Region space on the variables $\left(  \kappa,\sigma\right)  $ where the
points $P_{6}^{\pm}$ and $P_{7}^{\pm}$ are real. Second Column: Contour plots
of the effective equation of state parameters at the stationary points
$P_{6}^{\pm}$ and $P_{7}^{\pm}$. We observe that the effective equation of
state parameter can take value less than minus one.}%
\label{fig1}%
\end{figure}

The stationary points $P=\left(  x\left(  P\right)  ,y\left(  P\right)
,z\left(  P\right)  ,u\left(  P\right)  ,\mu\left(  P\right)  \right)  $ of
the five-dimensional dynamical system (\ref{ch.02})-(\ref{ch.06}) which
satisfy the algebraic equation (\ref{ch.00}) are as follows:%

\[
P_{1}^{\pm}=\left(  \pm1,0,0,0,0\right)  ,
\]
where only the kinetic components of the scalar field $\phi$ contributes to
the cosmological fluid, resulting in, $\Omega_{\phi}\left(  P_{1}^{\pm
}\right)  =1$ and $\Omega_{\psi}\left(  P_{1}^{\pm}\right)  =0$. Furthermore,
at the stationary points $P_{1}^{\pm}$ the effective equation of state
parameter is given by $w_{tot}\left(  P_{1}^{\pm}\right)  =1$, indicating that
the effective cosmological fluid corresponds to stiff matter. To examine the
stability properties of the stationary points we derive the eigenvalues of the
linearized system by replacing $y=\sqrt{1-u^{2}-x^{2}+z^{2}}$. The eigenvalues
are $e_{1}\left(  P_{1}^{\pm}\right)  =\mp\frac{\sqrt{6}}{2}\kappa
~,~e_{2}\left(  P_{1}^{\pm}\right)  =\mp\frac{\sqrt{6}}{2}\kappa$
,~$e_{3}\left(  P_{1}^{\pm}\right)  =\frac{\sqrt{6}}{2}\left(  \sqrt{6}%
\pm\kappa\right)  $ and $e_{4}\left(  P_{1}^{\pm}\right)  =\sqrt{6}\left(
\sqrt{6}\pm\lambda\right)  $. For $-\sqrt{6}<\kappa<0$ $\ $and $\lambda
>-\sqrt{6}$, stationary point $P_{1}^{+}$ is always a source, otherwise it is
a saddle point. Similarly, for $0<\kappa<\sqrt{6}$ and $\lambda<\sqrt{6}$,
point $P_{1}^{-}$ is a source, otherwise is a saddle point.%

\begin{equation}
P_{2}=\left(  -\frac{\lambda}{\sqrt{6}},\sqrt{1-\frac{\lambda^{2}}{6}%
},0,0,0\right)  ,
\end{equation}
with $\Omega_{\phi}\left(  P_{2}\right)  =1,~\Omega_{\psi}\left(
P_{2}\right)  =0$ and $w_{tot}\left(  P_{2}\right)  =-1+\frac{\lambda^{2}}{3}%
$. The point exists in real space for $\lambda^{2}<6$ , and it represents the
scaling solution previously discovered for the quintessence scalar field
\cite{cop}. Acceleration occurs when $\lambda^{2}<2.$ The eigenvalues of the
linearized system are derived $e_{1}\left(  P_{2}\right)  =\frac{\kappa
\lambda}{2}$,~$e_{2}\left(  P_{2}\right)  =\frac{\lambda^{2}-6}{2}$%
,~$e_{3}\left(  P_{2}\right)  =\frac{\lambda}{2}\left(  \lambda-\kappa\right)
$ and $e_{4}\left(  P_{2}\right)  =\frac{1}{2}\left(  \lambda\left(
\lambda+\kappa\right)  -6\right)  $. Therefore, point $P_{2}$ is always a
saddle point.%

\begin{equation}
P_{3}^{\pm}=\left(  -\frac{\sqrt{6}}{\kappa+\lambda},\sqrt{\frac{\kappa
}{\kappa+\lambda}},\pm\frac{\sqrt{6-\lambda\left(  \kappa+\lambda\right)  }%
}{\kappa+\lambda},0,0\right)  ,
\end{equation}
with $\Omega_{\phi}\left(  P_{3}^{\pm}\right)  =\frac{\kappa\left(
\kappa+\lambda\right)  +6}{\left(  \kappa+\lambda\right)  ^{2}},~\Omega_{\psi
}\left(  P_{3}^{\pm}\right)  =\frac{\lambda\left(  \kappa+\lambda\right)
-6}{\left(  \kappa+\lambda\right)  ^{2}}$ and $w_{tot}\left(  P_{3}^{\pm
}\right)  =\frac{\lambda-\kappa}{\lambda+\kappa}$. Points $P_{3}^{\pm}$ are
real and physically accepted for $\left\{  \lambda<0,\kappa>-\lambda\right\}
$ or $\left\{  \lambda>0,\kappa<-\lambda\right\}  $ or $\left\{  \kappa
\lambda>0,\lambda\left(  \kappa+\lambda\right)  <6\right\}  $ or $\left\{
\lambda=0,\kappa\neq0\right\}  $. The asymptotic solutions at the stationary
points $P_{3}^{\pm}$ describes the hyperbolic inflation \cite{vr91} when
$\frac{\lambda-\kappa}{\lambda+\kappa}<-\frac{1}{3}$. We remark that
$w_{tot}\left(  P_{3}^{\pm}\right)  $ crosses the phantom divide line
$\left\{  \lambda>0,\kappa<-\lambda\right\}  $ and $\left\{  \lambda
<0,\kappa>-\lambda\right\}  .~$ The eigenvalues of the linearized system
are~$e_{1}\left(  P_{3}^{\pm}\right)  =\frac{3\kappa}{\kappa+\lambda}%
$~,~$e_{2}\left(  P_{3}^{\pm}\right)  =3\left(  1-\frac{2\kappa}%
{\kappa+\lambda}\right)  $, $e_{3}\left(  P_{3}^{\pm}\right)  =-\frac{3}%
{2}\frac{\kappa}{\kappa+\lambda}+i\frac{\sqrt{3\kappa\left(  4\lambda
^{2}\left(  \lambda+2\kappa\right)  +4\lambda\left(  \kappa^{2}-6\right)
-27\kappa\right)  }}{2\left(  \kappa+\lambda\right)  }$ and $e_{4}\left(
P_{3}^{\pm}\right)  =-\frac{3}{2}\frac{\kappa}{\kappa+\lambda}-i\frac
{\sqrt{3\kappa\left(  4\lambda^{2}\left(  \lambda+2\kappa\right)
+4\lambda\left(  \kappa^{2}-6\right)  -27\kappa\right)  }}{2\left(
\kappa+\lambda\right)  }$. We conclude that the stationary points are always
saddle points.%

\begin{equation}
P_{4}^{\pm}=\left(  -\frac{\sqrt{6}}{\kappa+\lambda},\sqrt{\frac{\kappa
}{\kappa+\lambda}},\pm\frac{\sqrt{6-\lambda\left(  \kappa+\lambda\right)  }%
}{\kappa+\lambda},0,\pm\frac{\sqrt{6}\kappa}{\sigma\sqrt{6-\lambda\left(
\kappa+\lambda\right)  }}\right)  ,
\end{equation}
have the same physical properties and existence conditions with points
$P_{3}^{\pm}$. Indeed, the stationary points $P_{4}^{\pm}$ describe the
hyperbolic inflation for $\frac{\lambda-\kappa}{\lambda+\kappa}<-\frac{1}{3}$.
The eigenvalues of the linearized system are derived $e_{1}\left(  P_{4}^{\pm
}\right)  =-\frac{3\kappa}{\kappa+\lambda}$, $e_{2}\left(  P_{4}^{\pm}\right)
=\frac{3\left(  \kappa+2\sigma\left(  \lambda-\sigma\right)  \right)
}{2\sigma\left(  \kappa+\lambda\right)  }$,$~e_{3}\left(  P_{4}^{\pm}\right)
=-\frac{3\kappa}{2\left(  \kappa+\lambda\right)  }+i\frac{\sqrt{3\kappa\left(
4\lambda^{2}\left(  \lambda+2\kappa\right)  +4\lambda\left(  \kappa
^{2}-6\right)  -27\kappa\right)  }}{2\left(  \kappa+\lambda\right)  }$ and
$e_{4}\left(  P_{4}^{\pm}\right)  =-\frac{3\kappa}{2\left(  \kappa
+\lambda\right)  }-i\frac{\sqrt{3\kappa\left(  4\lambda^{2}\left(
\lambda+2\kappa\right)  +4\lambda\left(  \kappa^{2}-6\right)  -27\kappa
\right)  }}{2\left(  \kappa+\lambda\right)  }$. Hence, points $P_{4}^{\pm}$
are always saddle points.%

\[
P_{5}^{\pm}=\left(  -\frac{\sqrt{6}}{2\kappa},0,\pm\frac{\sqrt{6-2\kappa^{2}}%
}{2\kappa},\frac{\sqrt{2}}{2},0\right)
\]
with $\Omega_{\phi}\left(  P_{5}^{\pm}\right)  =\frac{3}{2\kappa^{2}}$,
$\Omega_{\psi}\left(  P_{5}^{\pm}\right)  =1-\frac{3}{2\kappa^{2}}$ and
$w_{tot}\left(  P_{5}^{\pm}\right)  =0$. As a result, the asymptotic solutions
at the stationary points $P_{5}^{\pm}$ describe a universe where the
cosmological fluid is a dust fluid. The exact solution of the scale factor is
$a\left(  t\right)  =a_{0}t^{\frac{2}{3}}$, indicating that these points
describe the matter-dominated era in the cosmological evolution. The
eigenvalues of the linearized system are computed, $e_{1}\left(  P_{5}^{\pm
}\right)  =\frac{3}{2}$, $e_{2}\left(  P_{5}^{\pm}\right)  =3\left(
1-\frac{\lambda}{\kappa}\right)  $,~$e_{3}\left(  P_{5}^{\pm}\right)
=-\frac{3}{4}+\frac{\sqrt{3\left(  51-16\kappa^{2}\right)  }}{4}$ and
$e_{4}\left(  P_{5}^{\pm}\right)  =-\frac{3}{4}-\frac{\sqrt{3\left(
51-16\kappa^{2}\right)  }}{4}$. Consequently, the eigenvalues reveals that
points $P_{5}^{\pm}$ are always saddle points.
\[
P_{6}^{\pm}=\left(  x_{6},0,\pm z_{6},\sqrt{1+x_{6}^{2}-z_{6}^{2}+\frac
{\kappa\left(  2\sigma-1\right)  }{\sqrt{6}\sigma}x_{6}},-\frac{\kappa x_{6}%
}{2\sigma z_{6}}\right)  ,
\]
where $z_{6}=\sqrt{\frac{-\kappa x_{6}\left(  \sqrt{6}\sigma+\kappa\left(
1+2\sigma\right)  x_{6}+\sqrt{6}\sigma x_{6}^{2}\right)  }{2\sigma\left(
6\sigma+\sqrt{6}\kappa\left(  2\sigma-1\right)  x_{6}\right)  }}$ and
$x_{6}=\frac{\kappa^{2}\left(  1-2\sigma\right)  -6\sigma+\sqrt{\kappa
^{4}\left(  1-4\sigma\right)  +4\sigma^{2}\left(  \kappa^{2}-3\right)  ^{2}}%
}{\sqrt{6}\kappa\left(  4\sigma-1\right)  }$. We derive $\Omega_{\phi}\left(
P_{6}^{\pm}\right)  =\frac{\left(  \kappa^{2}\left(  1-2\sigma\right)
-6\sigma+\sqrt{\kappa^{4}\left(  1-4\sigma\right)  +4\sigma^{2}\left(
\kappa^{2}-3\right)  ^{2}}\right)  }{6\kappa^{2}\left(  1-4\sigma\right)
^{2}}^{2}$ and $w_{tot}\left(  P_{6}^{\pm}\right)  =\frac{\kappa^{2}\left(
1-2\sigma\right)  ^{2}-12\sigma^{2}+\left(  1-2\sigma\right)  \sqrt{\kappa
^{4}\left(  1-4\sigma\right)  +4\sigma^{2}\left(  \kappa^{2}-3\right)  ^{2}}%
}{6\sigma\left(  4\sigma-1\right)  }$. In Fig. \ref{fig1} we present the
region in the two-dimensional space $\left(  \kappa,\sigma\right)  $ where
points $P_{6}^{\pm}$ are real, and the contour plot for $w_{tot}\left(
P_{6}^{\pm}\right)  $ from where we see that $w_{tot}\left(  P_{6}^{\pm
}\right)  $ can take values less than the minus one.%

\[
P_{7}^{\pm}=\left(  x_{7},0,\pm z_{7},\sqrt{1+x_{7}^{2}-z_{7}^{2}+\frac
{\kappa\left(  2\sigma-1\right)  }{\sqrt{6}\sigma}x_{7}},-\frac{\kappa x_{7}%
}{2\sigma z_{7}}\right)  ,
\]
where $z_{7}=\sqrt{\frac{-\kappa x_{7}\left(  \sqrt{6}\sigma+\kappa\left(
1+2\sigma\right)  x_{7}+\sqrt{6}\sigma x_{7}^{2}\right)  }{2\sigma\left(
6\sigma+\sqrt{6}\kappa\left(  2\sigma-1\right)  x_{7}\right)  }}$ and
$x_{7}=\frac{\kappa^{2}\left(  1-2\sigma\right)  -6\sigma-\sqrt{\kappa
^{4}\left(  1-4\sigma\right)  +4\sigma^{2}\left(  \kappa^{2}-3\right)  ^{2}}%
}{\sqrt{6}\kappa\left(  4\sigma-1\right)  }$. Moreover, we calculate
$\Omega_{\phi}\left(  P_{7}^{\pm}\right)  =\frac{\left(  \kappa^{2}\left(
1-2\sigma\right)  +6\sigma+\sqrt{\kappa^{4}\left(  1-4\sigma\right)
+4\sigma^{2}\left(  \kappa^{2}-3\right)  ^{2}}\right)  }{6\kappa^{2}\left(
1-4\sigma\right)  ^{2}}^{2}$ and $w_{tot}\left(  P_{7}^{\pm}\right)
=\frac{\kappa^{2}\left(  1-2\sigma\right)  ^{2}+12\sigma^{2}-\left(
1-2\sigma\right)  \sqrt{\kappa^{4}\left(  1-4\sigma\right)  +4\sigma
^{2}\left(  \kappa^{2}-3\right)  ^{2}}}{6\sigma\left(  4\sigma-1\right)  }$.
In Fig. \ref{fig1} we present the region in the two-dimensional space $\left(
\kappa,\sigma\right)  $ where points $P_{7}^{\pm}$ are real, and the contour
plot for $w_{tot}\left(  P_{6}^{\pm}\right)  $ from where we see that
$w_{tot}\left(  P_{7}^{\pm}\right)  $ can take values smaller than the minus one.

Due to the intricate nature of the eigenvalues of the linearized system around
the points $P_{6}^{\pm}$ and $P_{7}^{\pm}$ we conducted a numerical analysis
to investigate their stability properties. By employing random numbers for
parameters $\lambda,~\kappa$ and $\sigma$ we performed multiple runs. Our
findings consistently indicated that the stationary points $P_{6}^{\pm}$,
$P_{7}^{\pm}$ are saddle points.%

\[
P_{8}=\left(  -\frac{\kappa}{\sqrt{6}},0,0,\sqrt{1-\frac{\kappa^{2}}{6}%
},0\right)
\]
which a scaling solution with $w_{tot}\left(  P_{8}\right)  =-1+\frac
{\kappa^{2}}{3}$. The point is real and physically accepted for $\kappa^{2}%
<6$. We remark that point $P_{8}$ has similarities with $P_{2}$ where now the
exponential term in the mixed potential drive the dynamics and the
second-scalar field is constant. \ The eigenvalues of the linearized system
are $e_{1}\left(  P_{8}\right)  =\frac{\kappa^{2}}{2}$,~$e_{2}\left(
P_{8}\right)  =\frac{\kappa^{2}-6}{2}$, $e_{3}\left(  P_{8}\right)
=\kappa^{2}-3$ and $e_{4}\left(  P_{8}\right)  =\kappa\left(  \kappa
-\lambda\right)  $, which means that $P_{8}$ is always a saddle point.%

\[
P_{9}=\left(  0,\sqrt{\frac{\kappa}{\kappa-\lambda}},0,\sqrt{\frac{\lambda
}{\kappa-\lambda}},0\right)  ,
\]
describes a de Sitter universe with $w_{tot}\left(  P_{9}\right)  =-1$. The
point is real for $\left\{  \lambda<0,\kappa<\lambda\right\}  $ or $\left\{
\lambda>0,\kappa>\lambda\right\}  $. The eigenvalues of the linearized system
are $e_{1}\left(  P_{9}\right)  =0$, $e_{2}\left(  P_{9}\right)  =-3$,
$e_{3}\left(  P_{9}\right)  =\frac{1}{2}\left(  -3+\sqrt{3\left(
3+4\kappa\lambda\right)  }\right)  $ and $e_{4}\left(  P_{9}\right)  =\frac
{1}{2}\left(  -3-\sqrt{3\left(  3+4\kappa\lambda\right)  }\right)  $, which
means that $P_{9}$ is always a saddle point.

The above results are summarized in Table \ref{tab1}.%

\begin{table}[tbp] \centering
\caption{Stationary points at the finite regime for the Chiral-Quintom model with a mixed potential}%
\begin{tabular}
[c]{lllll}\hline\hline
\textbf{Point} & $\mathbf{\Omega}_{\psi}$ & \textbf{Acceleration} &
$\mathbf{w}_{tot}<-1$ & \textbf{Stability}\\\hline
$P_{1}^{\pm}$ & $0$ & No $\left(  w_{tot}=1\right)  $ & No & Saddle\\
$P_{2}$ & $0$ & Yes & No & Saddle\\
$P_{3}^{\pm}$ & $\neq0$ & Yes & Yes & Saddle\\
$P_{4}^{\pm}$ & $\neq0$ & Yes & Yes & Saddle\\
$P_{5}^{\pm}$ & $\neq0$ & No $\left(  w_{tot}=0\right)  $ & No & Saddle\\
$P_{6}^{\pm}$ & $\neq0$ & Yes & Yes & Saddle\\
$P_{7}^{\pm}$ & $\neq0$ & Yes & Yes & Saddle\\
$P_{8}$ & $\neq0$ & Yes & No & Saddle\\
$P_{9}$ & $\neq0$ & Yes~$\left(  w_{tot}=-1\right)  $ & No &
Saddle\\\hline\hline
\end{tabular}
\label{tab1}%
\end{table}%

\subsection{Stationary points at the infinity}

We proceed with the definition of the Poincare variables
\[
x=\frac{X}{\rho}~,~z=\frac{Z}{\rho}\text{~},~u=\frac{U}{\rho}%
\]
where $\rho=\sqrt{1-X^{2}-Z^{2}-U^{2}}~$and we have used the constraint
equation $y=\sqrt{1-x^{2}+z^{2}-u^{2}}$.

In the new variables $\left(  X,Z,U\,\right)  $ the field equations read%
\begin{equation}
\frac{dX}{dT}=-\frac{1}{2}\left(  1-2X^{2}\right)  \left(  6X\rho+\sqrt
{6}\left(  \lambda\left(  1-X^{2}\right)  +\kappa Z^{2}\right)  \right)
-\frac{\sqrt{6}}{2}\left(  \kappa-2\lambda+2X\left(  \lambda X+\mu Z\right)
\right)  ~, \label{c.01}%
\end{equation}%
\begin{equation}
\frac{dZ}{dT}=\frac{\sqrt{6}}{2}\lambda XZ\left(  1-2\left(  X^{2}%
+U^{2}\right)  \right)  -3Z\rho\left(  1-2X^{2}\right)  -\frac{\sqrt{6}}%
{2}\left(  1-2Z^{2}\right)  \left(  \kappa XZ-\mu U^{2}\right)  ~,
\label{c.02}%
\end{equation}%
\begin{equation}
\frac{dU}{dT}=6UX^{2}\rho-\sqrt{6}U^{3}\left(  \lambda X+\mu Z\right)
+\frac{\sqrt{6}}{2}U\left(  X\left(  \kappa\left(  1+2Z^{2}\right)
+\lambda\right)  +\mu Z-2\lambda X^{3}\right)  ~, \label{c.03}%
\end{equation}%
\begin{equation}
\frac{d\mu}{dT}=-\frac{\sqrt{6}}{2}\mu\left(  \kappa X+2\sigma\mu Z\right)  ~,
\label{c.04}%
\end{equation}
in which $T$ is a new independent variable $dT=\rho d\tau.$

The stationary points at the infinity are the points on the surface $\rho=0$.
At each stationary point the effective equation of state parameter for the
cosmological fluid is%
\begin{equation}
w_{tot}\left(  X,Z,U\right)  =-\frac{1-3X^{2}+Z^{2}-U^{2}}{\rho}. \label{c.05}%
\end{equation}

The stationary points~$Q=\left(  X\left(  Q\right)  ,Z\left(  P\right)
,U\left(  P\right)  ,\mu\left(  P\right)  \right)  $ of the dynamical system
(\ref{c.01})-(\ref{c.04}) at the infinity are%
\[
Q_{1}^{\pm}=\left(  \pm1,0,0,0\right)  ,
\]%
\[
Q_{2}^{\pm}=\left(  0,\pm\sqrt{\frac{\lambda-\kappa}{2\lambda}},\sqrt
{\frac{\kappa+\lambda}{2\lambda}}0\right)  ,
\]%
\[
Q_{3}^{\pm}=\left(  \frac{\sqrt{2}}{2},\pm\frac{\sqrt{2}}{2},0,0\right)  ,
\]%
\[
Q_{4}^{\pm}=\left(  -\frac{\sqrt{2}}{2},\pm\frac{\sqrt{2}}{2},0,0\right)  ,
\]%
\[
Q_{5}^{\pm}=\left(  \frac{\sqrt{2}}{2},\pm\frac{\sqrt{2}}{2},0,\mp\frac
{\kappa}{2\sigma}\right)  ,
\]%
\[
Q_{6}^{\pm}=\left(  -\frac{\sqrt{2}}{2},\pm\frac{\sqrt{2}}{2},0,\pm
\frac{\kappa}{2\sigma}\right)  ,
\]%
\[
Q_{7}^{\pm}=\left(  \sqrt{2\sigma},\pm\frac{\sqrt{2}}{2},\sqrt{\frac
{1-4\sigma}{2}},\mp\frac{\kappa}{\sqrt{6}}\right)  ,
\]%
\[
Q_{8}^{\pm}=\left(  -\sqrt{2\sigma},\pm\frac{\sqrt{2}}{2},\sqrt{\frac
{1-4\sigma}{2}},\pm\frac{\kappa}{\sqrt{6}}\right)  .
\]

The existence conditions indicate that, points $Q_{1}^{\pm}$ are not
physically accepted and for the points $Q_{2}^{\pm}$ it follows~$\left\{
\lambda<0,~0<\kappa<-\lambda\right\}  $, $\left\{  \lambda>0,-\lambda
<\kappa<0\right\}  $. Moreover, points $Q_{7}^{\pm}$ and $Q_{8}^{\pm}$ are
real for $\sigma>0$ and $1-4\sigma\geq0$, that is, $0<\sigma\leq\frac{1}{4}\,$.

By replacing in (\ref{c.05}) it follows that the stationary points $Q_{2}%
^{\pm},~Q_{7}^{\pm}$ and $Q_{8}^{\pm}$ describe a cosmological solution with
$w_{tot}\rightarrow-\infty$, that is Big Rip. However, for $\sigma=\frac{1}%
{4}$ the stationary points $Q_{7}^{\pm}$ and $Q_{8}^{\pm}$ describe dust fluid
solutions with $w_{tot}\left(  Q_{7}^{\pm}\right)  =0$ and $w_{tot}\left(
Q_{8}^{\pm}\right)  =0$.~However, as we reach the limit of the rest of the
stationary points we derive $w_{tot}\left(  Q_{3,4,5,6}^{\pm}\right)  =0$,
that is, these points describe dust fluid asymptotic solutions.

Regarding the stability properties of the stationary points, we omit the
presentation of the analysis but we conclude that the stationary points at the
infinity when they exist are always saddle points.

\begin{figure}[ptbh]
\centering\includegraphics[width=1\textwidth]{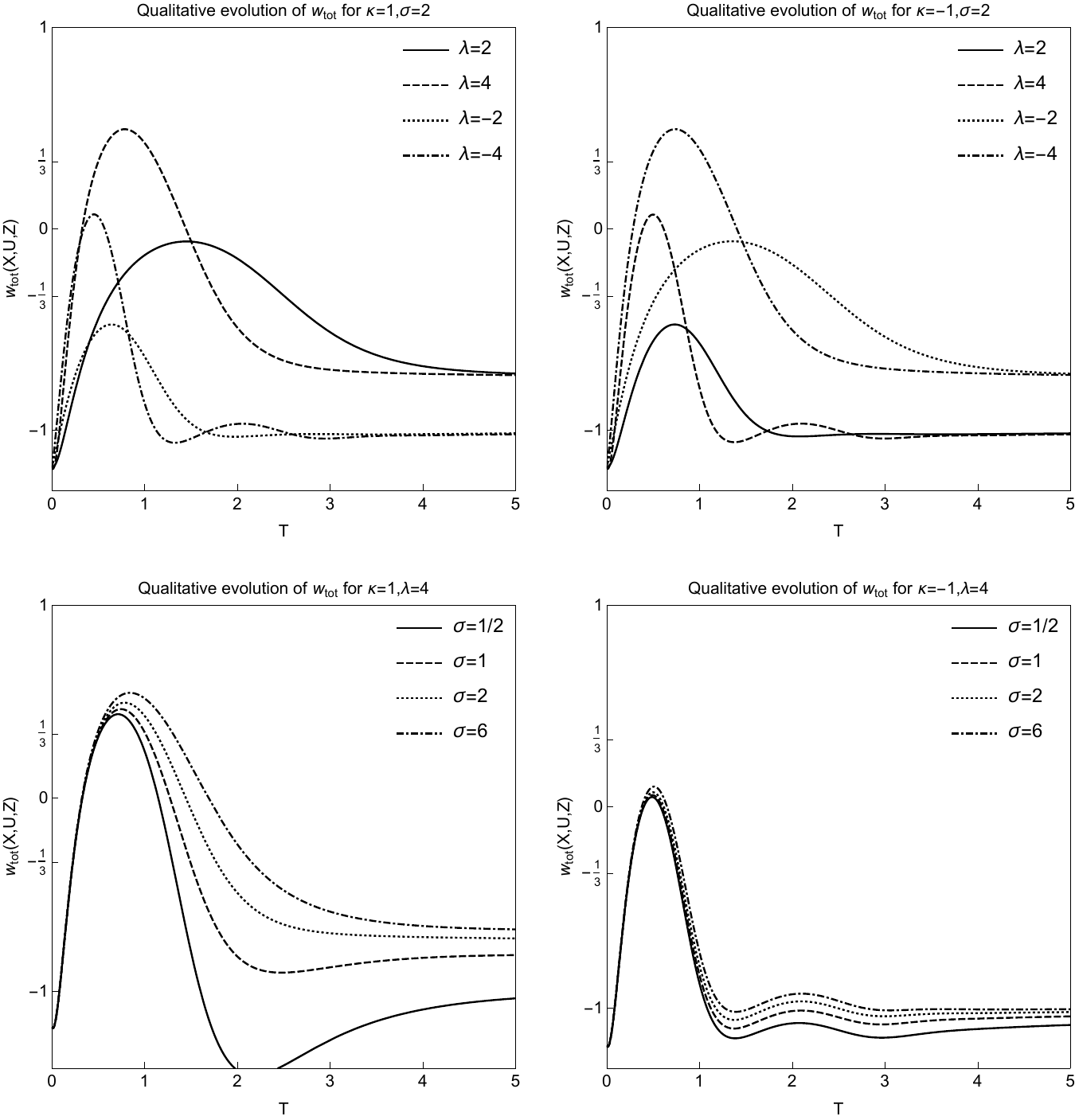}\caption{Qualitative
evolution of the equation of state parameter (\ref{c.05}) as it is given by
numerical simulations of the dynamical system  (\ref{c.01})-(\ref{c.04}) for
various values of the free parameters. For the plots we considered the initial
conditions $X_{0}=0.1,~Z_{0}=0.3$, $U_{0}=0.2$ and $\mu_{0}=2$. }%
\label{plot1}%
\end{figure}

\begin{figure}[ptbh]
\centering\includegraphics[width=1\textwidth]{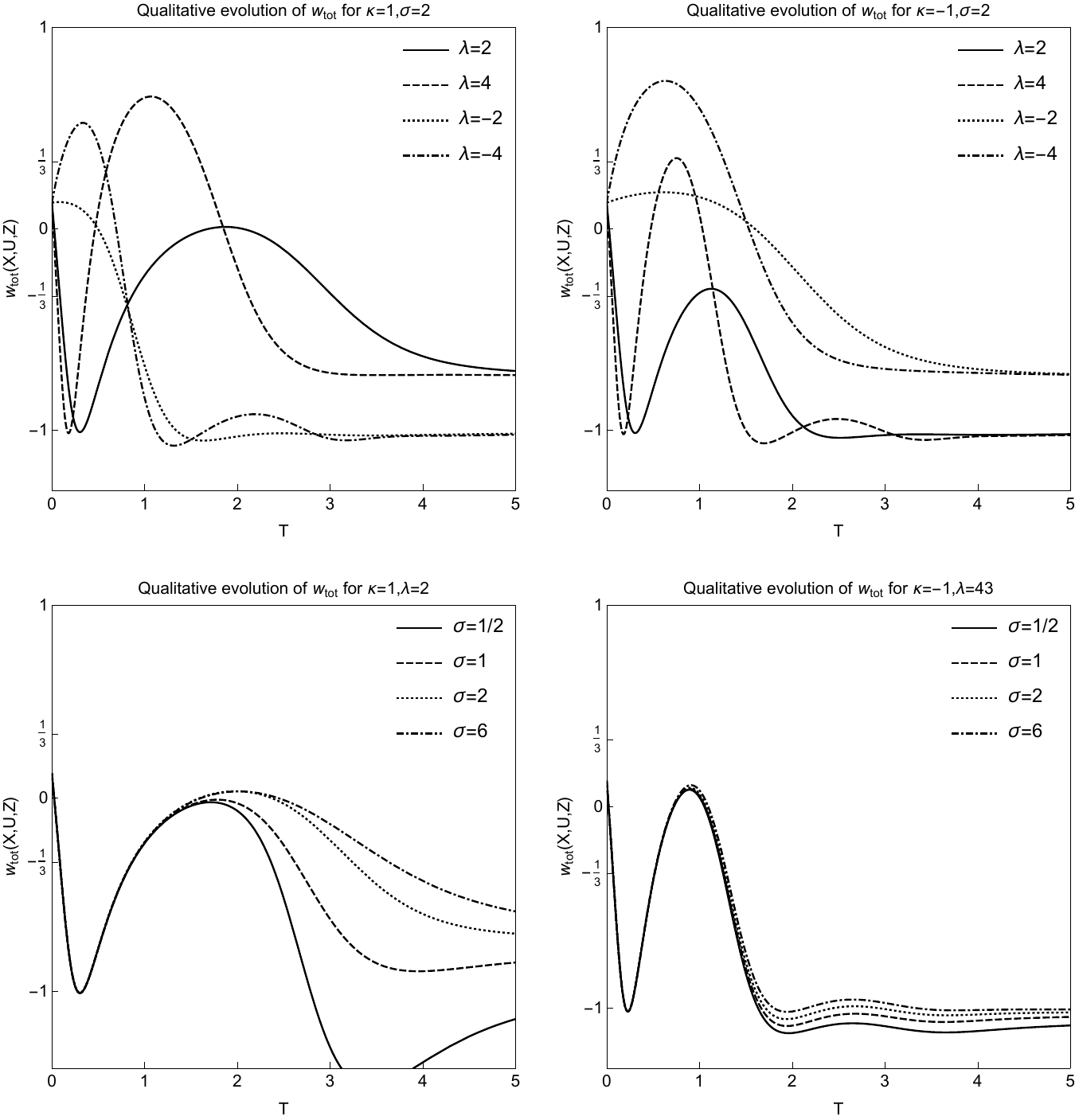}\caption{Qualitative
evolution of the equation of state parameter (\ref{c.05}) as it is given by
numerical simulations of the dynamical system (\ref{c.01})-(\ref{c.04}) for
various values of the free parameters. For the plots we considered the initial
conditions $X_{0}=0.6,~Z_{0}=0.1$, $U_{0}=0.1$ and $\mu_{0}=2$. }%
\label{plot2}%
\end{figure}

In Figs. \ref{plot1} and \ref{plot2} we present the qualitative evolution of
the effective equation of state parameter $w_{tot}\left(  X,Z,U\right)  $ as
it is given by the numerical solution of the dynamical system (\ref{c.01}%
)-(\ref{c.04}). The Figs are for different sets of initial conditions and
different values of the free parameters. We remark that this scalar field
model can describe a unification of the dark matter and of the dark energy in
the cosmic evolution.

\section{Asymptotic solutions for potential $V\left(  \phi,\psi\right)
=U_{0}e^{\kappa\phi}\psi^{\frac{1}{\sigma}}$}

\label{sec4a}

Consider the \ second scenario where the mixed potential function takes the
form $V\left(  \phi,\psi\right)  =U_{0}e^{\kappa\phi}\psi^{\frac{1}{\sigma}}$,
which means that the dynamical system is defined on the surface with $y=0$.

\subsection{Stationary points at the finite regime}

Assume the potential function $V\left(  \phi,\psi\right)  =U_{0}e^{\kappa\phi
}\psi^{\frac{1}{\sigma}}$, where $y$ is always zero. As a result, the
dynamical system's dimension is decreased by one. The stationary points for
this potential are the one derived before with $y=0$.

Points $A=\left(  x\left(  A\right)  ,z\left(  A\right)  ,u\left(  A\right)
,\mu\left(  A\right)  \right)  $ of the dynamical system on the surface
$1-x^{2}+z^{2}-u^{2}=0$, are:
\[
A_{1}^{\pm}=\left(  \pm1,0,0,0\right)  ,
\]

\[
A_{2}^{\pm}=\left(  -\frac{\sqrt{6}}{2\kappa},\pm\frac{\sqrt{6-2\kappa^{2}}%
}{2\kappa},\frac{\sqrt{2}}{2},0\right)
\]

\[
A_{3}^{\pm}=\left(  x_{6},\pm z_{6},\sqrt{1+x_{6}^{2}-z_{6}^{2}+\frac
{\kappa\left(  2\sigma-1\right)  }{\sqrt{6}\sigma}x_{6}},-\frac{\kappa x_{6}%
}{2\sigma z_{6}}\right)  ,
\]%
\[
A_{4}^{\pm}=\left(  x_{7},\pm z_{7},\sqrt{1+x_{7}^{2}-z_{7}^{2}+\frac
{\kappa\left(  2\sigma-1\right)  }{\sqrt{6}\sigma}x_{7}},-\frac{\kappa x_{7}%
}{2\sigma z_{7}}\right)  ,
\]%
\[
A_{5}=\left(  -\frac{\kappa}{\sqrt{6}},0,\sqrt{1-\frac{\kappa^{2}}{6}%
},0\right)  .
\]
where $z_{6},$ $z_{7}~$, $x_{6}$ and $x_{7}$ are that of points $P_{6}^{\pm}$
and $P_{7}^{\pm}$ respectively.

The characteristics of the asymptotic solutions at these points are similar
with that found before. We recall that because $y=0$, the stationary points
which describe the hyperbolic inflation do not exist. However, inflation can
occur by the stationary points $A_{3}^{\pm}$ , $A_{4}^{\pm}$ and $A_{5}$.
Points $A_{2}^{\pm}$ describe universes dominated by a pressureless fluid source.

Because the dynamical system for this specific potential function lies on the
surface $y=0$,the stability properties of the stationary points may exhibit
variations. In particular, we observe that the stability properties of points
$A_{1}^{\pm},~A_{2}^{\pm}$ and$~A_{5}$ are the same as their related points
$P$, indicating that they are all saddle points. Nevertheless, the stability
property for the points $A_{3}^{\pm}$ and $A_{4}^{\pm}$ differs from the
others. Upon conducting a similar analysis as before, we determine that points
$A_{3}^{\pm}$ and $A_{4}^{\pm}$ can be attractors, exhibiting distinct
stability behavior compared to the aforementioned saddle points.

\begin{figure}[ptb]
\centering\includegraphics[width=1\textwidth]{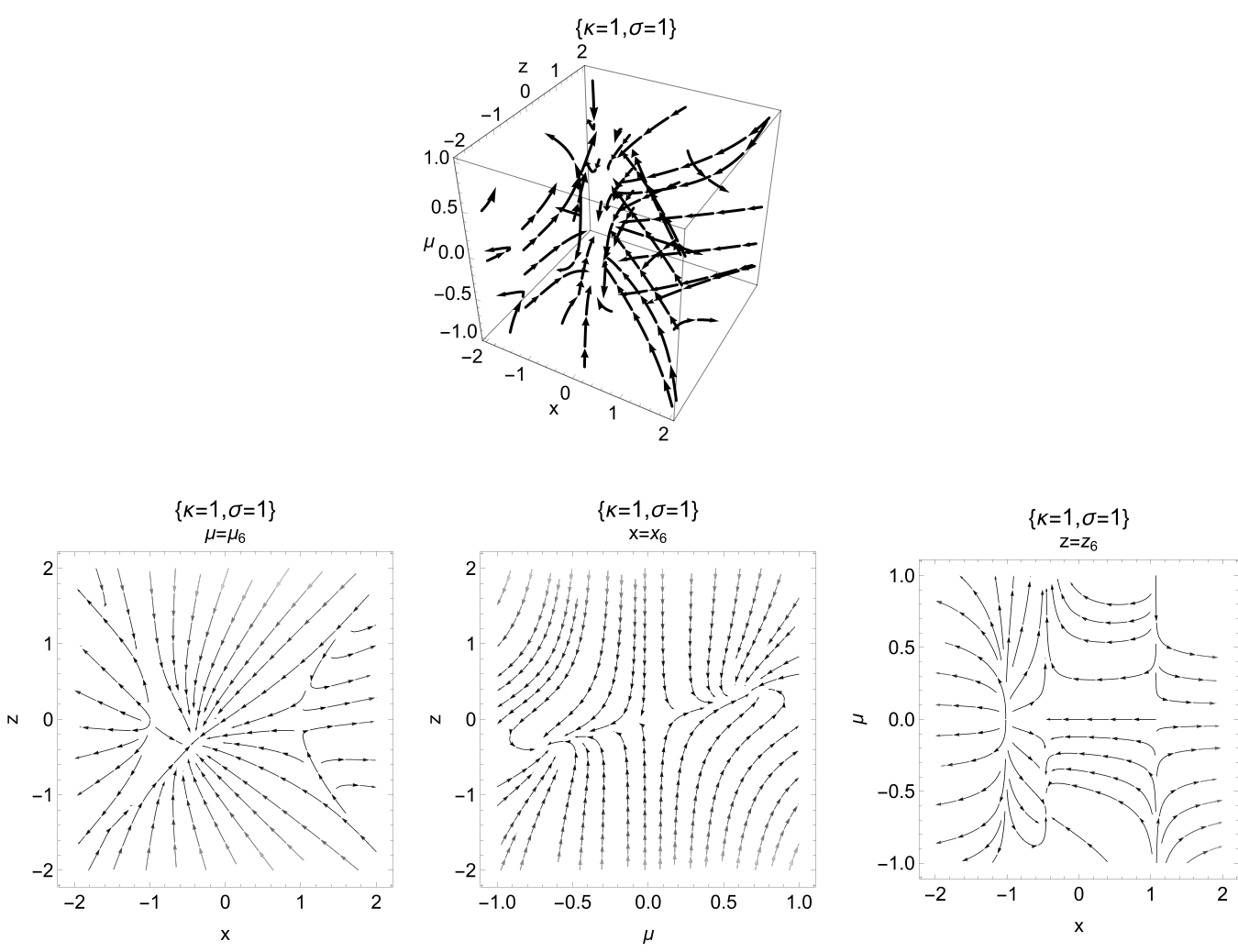}\caption{Phase-space
portrait for the dynamical system with the mixed potential function $V\left(
\phi,\psi\right)  =U_{0}e^{\kappa\phi}\psi^{\frac{1}{\sigma}}$. We observe
that the stationary points $A_{3}^{-}$ is an attractor.}%
\label{fig2}%
\end{figure}

In order to give a comprehensive view of the dynamics and behavior of the for
the field equations, in Fig. \ref{fig2} we demonstrate the phase-space
portraits for or $\kappa=1$ and $\sigma=1$. In particular we give a plot in
the three-dimensional space and on two-dimensional surfaces on the the values
where point $A_{3}^{-}$ is an attractor.\ For $\kappa=1$ and $\sigma=1$ it
holds $w_{tot}\left(  A_{3}^{\pm}\right)  \simeq-0.81$ which means that an
accelerated universe is described by point $A_{3}^{-}$. 

We conclude that this model can describe a cosmological history, with an early
acceleration phase (point $A_{5}$), a matter era (points $A_{2}^{\pm}$) and a
future acceleration point (points $A_{3}^{\pm}$ or $A_{4}^{\pm}$).

To ensure a thorough analysis, it is imperative to study the existence of
stationary points at the infinity regime.

\subsection{Stationary points at the infinity}

We introduce the new set of Poincare variables%
\[
x=\frac{X}{\bar{\rho}}~,~z=\frac{Z}{\bar{\rho}}\text{~},~d\bar{T}=\bar{\rho
}d\tau.
\]
where now $\rho=\sqrt{1-X^{2}-Z^{2}}~$and we have used the constraint
condition $u=\sqrt{1-x^{2}+z^{2}}$.

Therefore, the field equations in the Poincare variables $\left(  X,Z\right)
$ read%
\begin{equation}
\frac{dX}{d\bar{T}}=\frac{1}{2}\left(  1-2X^{2}\right)  \left(  \sqrt{6}%
\kappa\left(  X^{2}-1-Z^{2}\right)  -6X\bar{\rho}-\sqrt{6}XZ\mu\right)
,\label{d.01}%
\end{equation}%
\begin{equation}
\frac{dZ}{d\bar{T}}=3Z\bar{\rho}\left(  2X^{2}-1\right)  +\sqrt{6}%
\kappa\left(  Z^{2}-X^{2}\right)  +\frac{\sqrt{6}}{2}\mu\left(  1-2X^{2}%
\right)  \left(  1-Z^{2}\right)  ,
\end{equation}
and%
\begin{equation}
\frac{d\mu}{d\bar{T}}=-\frac{\sqrt{6}}{2}\mu\left(  \kappa X+2\sigma\mu
Z\right)  .\label{d.03}%
\end{equation}

The stationary points of the aforementioned dynamical system are of the form
$B=\left(  X\left(  B\right)  ,Z\left(  B\right)  ,\mu\left(  B\right)
\right)  $; they are
\[
B_{1}^{\pm}=\left(  \pm1,0,0\right)  ~,
\]%
\[
B_{2}^{\pm}=\left(  \frac{\sqrt{2}}{2},\pm\frac{\sqrt{2}}{2},0\right)  ~,
\]%
\[
B_{3}^{\pm}=\left(  -\frac{\sqrt{2}}{2},\pm\frac{\sqrt{2}}{2},0\right)  ~,
\]%
\[
B_{4}^{\pm}=\left(  \frac{\sqrt{2}}{2},\pm\frac{\sqrt{2}}{2},\mp\frac{\kappa
}{2\sigma}\right)  ,
\]%
\[
B_{5}^{\pm}=\left(  -\frac{\sqrt{2}}{2},\pm\frac{\sqrt{2}}{2},\pm\frac{\kappa
}{2\sigma}\right)  ,
\]%
\[
B_{6}^{\pm}=\left(  \sqrt{\frac{4\sigma}{1+4\sigma}},\pm\sqrt{\frac
{1}{1+4\sigma}},\mp\frac{\kappa}{\sqrt{\sigma}}\right)  ~,
\]%
\[
B_{7}^{\pm}=\left(  -\sqrt{\frac{4\sigma}{1+4\sigma}},\pm\sqrt{\frac
{1}{1+4\sigma}},\pm\frac{\kappa}{\sqrt{\sigma}}\right)  ~.
\]

Points $B_{1}^{\pm}$ are not physically accepted, while points $B_{6}^{\pm}$
and $B_{7}^{\pm}$ are real for~$0<\sigma<\frac{1}{4}\,$.

Regarding the physical properties of the stationary points $B_{2}^{\pm}$,
$B_{3}^{\pm}$, $B_{4}^{\pm}$ and $B_{5}^{\pm},$ the points describe dust fluid
asymptotic solutions; on the other hand, points $B_{6}^{\pm}$ and $B_{7}^{\pm
}$ correspond to Big Rip singularities.

The eigenvalues of the linearized system around the stationary points
$B_{2}^{\pm}$ and $B_{3}^{\pm}$ are $e_{1}\left(  B_{2}^{\pm}\right)
=-\frac{\sqrt{3}}{2}\kappa,~e_{2}\left(  B_{2}^{\pm}\right)  =\sqrt{3}\kappa
~$,~$e_{3}\left(  B_{2}^{\pm}\right)  =2\sqrt{3}\kappa$ and $e_{1}\left(
B_{3}^{\pm}\right)  =\frac{\sqrt{3}}{2}\kappa,~e_{2}\left(  B_{3}^{\pm
}\right)  =-\sqrt{3}\kappa~$,~$e_{3}\left(  B_{3}^{\pm}\right)  =-2\sqrt
{3}\kappa$ respectively. Hence, these two sets of points are always saddle
points. Furthermore, for the points $B_{4}^{\pm}$ and $B_{5}^{\pm}$ we derive
the eigenvalues $e_{1}\left(  B_{4}^{\pm}\right)  =-\frac{\sqrt{3}%
\kappa\left(  1-4\sigma\right)  }{2\sigma}\kappa,~e_{2}\left(  B_{4}^{\pm
}\right)  =\sqrt{3}\kappa~$,~$e_{3}\left(  B_{4}^{\pm}\right)  =2\sqrt
{3}\kappa$ and $e_{1}\left(  B_{5}^{\pm}\right)  =\frac{\sqrt{3}\kappa\left(
1-4\sigma\right)  }{2\sigma}\kappa,~e_{2}\left(  B_{5}^{\pm}\right)
=-\sqrt{3}\kappa~$,~$e_{3}\left(  B_{5}^{\pm}\right)  =-2\sqrt{3}\kappa$.
Therefore, points $B_{4}^{\pm}$ are attractors for $\left\{  \kappa
<0,\sigma<0\text{,}\sigma>\frac{1}{4}\right\}  $ while points $B_{5}^{\pm}$
are attractors for $\left\{  \kappa>0,0<\sigma<\frac{1}{4}\right\}  $.

Finally, the stability properties for the points $B_{6}^{\pm}$ and $B_{7}%
^{\pm}$ have been studied numerically. Based on our findings, we conclude that
the asymptotic solutions associated with these points are consistently unstable.

\begin{figure}[ptbh]
\centering\includegraphics[width=1\textwidth]{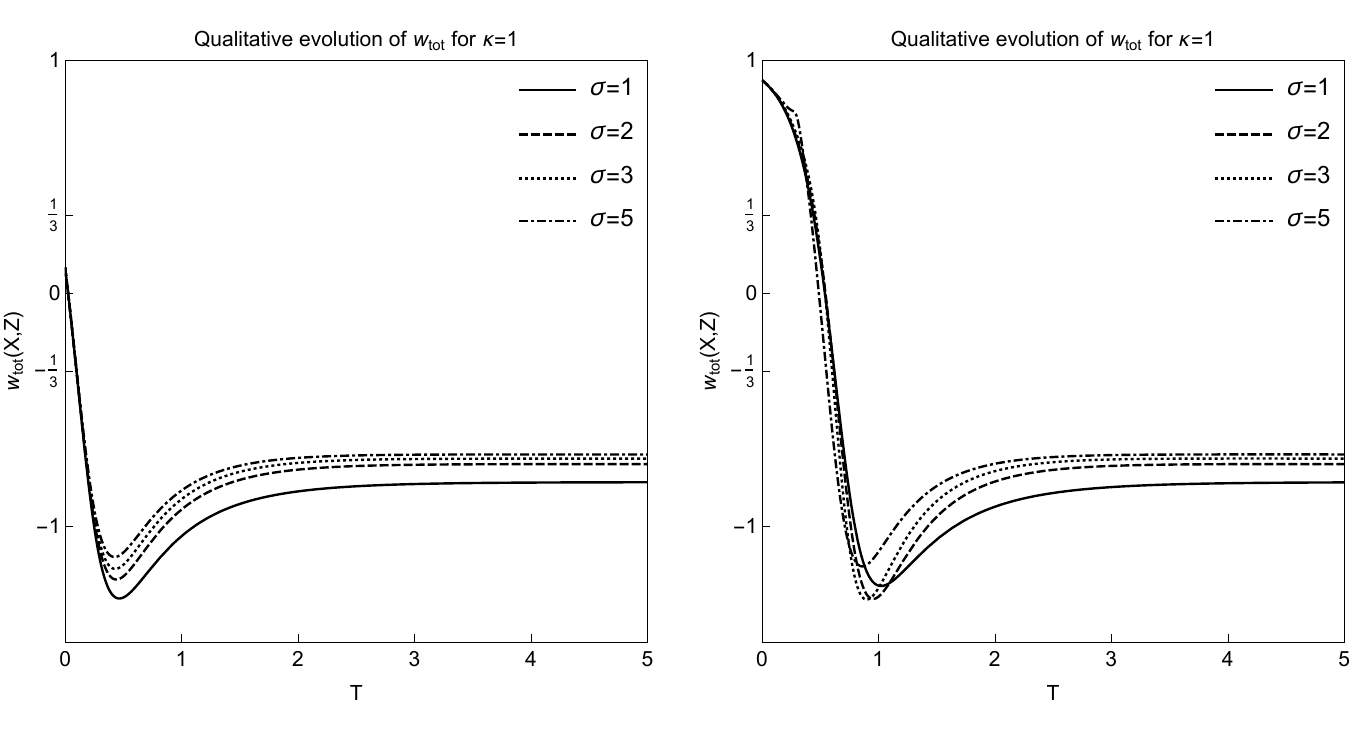}\caption{Qualitative
evolution of the equation of state parameter $w_{tot}\left(  X,Z\right)  $ as
it is given by numerical simulations of the dynamical system (\ref{d.01}%
)-(\ref{d.03}) for various values of the free parameters. Left Fig. is for the
initial conditions $\left(  X_{0},Z_{0},\mu_{0}\right)  =\left(
0.6,0.1,2\right)  $ while right Fig. is for the initial conditions $\left(
X_{0},Z_{0},\mu_{0}\right)  =\left(  0.7,0.2,-2\right)  $. }%
\label{plot3}%
\end{figure}

The qualitative evolution of the effective equation of state parameter for
this model is presented in Fig. \ref{plot3}.

\section{Conclusions}

\label{sec5}

The Chiral model is a multi-scalar field cosmological scenario which has been
proposed to describe inflation. In particular the inflationary mechanism
generated by the Chiral model is known as hyperbolic inflation. In this study
we considered the Chiral-quintom model which is a generalization where one of
the scalar fields has phantom energy component. As a result, the hyperbolic
inflationary mechanism is generalized where now the equation of state
parameter can cross the phantom divide line.

Considering a spatially flat FLRW geometry within this model, we introduced a
mixed potential term to modify the dynamics of the Chiral-quintom fluid.
Through a comprehensive analysis of the phase-space of the field equations, we
successfully reconstructed the complete cosmological history provided by this
model. Remarkably, this new multi-scalar field model effectively replicates
cosmological epochs that encompass the early-time and the late-time
acceleration phases of the universe as well as the matter-dominated epoch.
Consequently, this two-scalar field model holds promise as a unification
framework for the dark sector of the universe. We remark that the cosmological
history obtained from the same model without the mixed potential term
\cite{dn8} can be considered a special case of this more general model

In a forthcoming study, we plan to exploer the dynamical evolution of
perturbations within this multi-scalar field model featuring the mixed
potential. Additionally, we find it particularly intriguing to investigate
whether Chiral models can offer potential solutions to reconcile cosmological
tensions that exist in current observations and measurements.

\textbf{Data Availability Statements:} Data sharing is not applicable to this
article as no datasets were generated or analyzed during the current study.

\begin{acknowledgments}
AP was partially financially supported by the National Research Foundation of
South Africa (Grant Numbers 131604). AP thanks the support of
Vicerrector\'{\i}a de Investigaci\'{o}n y Desarrollo Tecnol\'{o}gico (Vridt)
at Universidad Cat\'{o}lica del Norte through N\'{u}cleo de Investigaci\'{o}n
Geometr\'{\i}a Diferencial y Aplicaciones, Resoluci\'{o}n Vridt No - 098/2022.
\end{acknowledgments}

\end{document}